\documentclass[preprint,12pt]{elsarticle}
\usepackage{amsmath}
\usepackage{epsfig,amssymb}

\newcommand{\sta}{{\rm sta}}

\newcommand{\calA}{{\cal{A}}}

\newcommand{ \PP}{{P}}

\newcommand{\vsig}{\varsigma}

\newcommand{\barx}{\bar{x}}

\newcommand{\real}{{\mathbb R}} 

\newcommand{\half}{\frac{1}{2}}

\newcommand{\bsig}{\mbox{\boldmath$\sigma$}}

\newcommand{\bvsig}{\mbox{\boldmath$\varsigma$}}

\newcommand{\veps}{\xi}

\newcommand{\bveps}{\mbox{\boldmath$\xi$}}

\newcommand{\barbsig}{\bar{\mbox{\boldmath$\sigma$}}}
\newcommand{\barbvsig}{\bar{\mbox{\boldmath$\varsigma$}}}

\newcommand{\beps}{\mbox{\boldmath$\epsilon$}}

\newcommand{\bxi}{{\mbox{\boldmath$\xi$}}}

\newcommand{\la}{\langle}
\newcommand{\ra}{\rangle}

\newcommand{\ba}{{{\bf a}}}

\newcommand{\bQ}{{\bf Q}}

\newcommand{\eba}{\begin{array}}
\newcommand{\eea}{\end{array}}
\newcommand{\ebe}{\begin{eqnarray}}
\newcommand{\eee}{\end{eqnarray}}
\newcommand{\eb}{\begin{equation}}
\newcommand{\ee}{\end{equation}}

\newcommand{\calW}{{\cal{W}}}
\newcommand{\calP}{{\cal{P}}}

\newcommand{\bC}{{\bf C}}
\newcommand{\bG}{{\bf G}}
\newcommand{\bH}{{\bf H}}

\newcommand{\bt}{{\bf t}}
\newcommand{\bff}{{\bf f}}

\newcommand{\bR}{{\bf R}}

\newcommand{\bw}{{\bf w}}

\newcommand{\bx}{{\bf x}}
\newcommand{\by}{{\bf y}}

\newcommand{\bX}{{\bf X}}

\newcommand{\bF}{{\bf F}}
\newcommand{\bA}{{\bf A}}
\newcommand{\bB}{{\bf B}}
\newcommand{\bI}{{\bf I}}

\newcommand{\calS}{{\cal S}}

\newcommand{\calE}{{\cal E}}

\newcommand{\calX}{{\cal X}}

\newcommand{\barbx}{\bar{\bf x}}
\newcommand{\barby}{\bar{\bf y}}

\newcommand{\bs}{{\bf s}}

\newcommand{\bd}{{\bf d}}

\newcommand{\be}{{\bf e}}

\newcommand{\alp}{{\alpha}}

\newcommand{ \eps}{{\epsilon}}

\newcommand{ \sig}{{\sigma}}
\newcommand{ \Lam}{{\Lambda}}

\newcommand{ \lam}{{\lambda}}

\newcommand{ \xx}{{\bf x}}

\newcommand{\bD}{{\bf D}}

\newcommand{\Diag}{{\mbox{Diag }}}

\newtheorem{definition}{Definition}

\newcommand{\bdelta}{\mbox{\boldmath$\delta$}}

\newcommand{\calXa}{{\calX_a}}
\newcommand{\FF}{{F}}
\newcommand{\VV}{{V}}

\newcommand{\WW}{{W}}

\newcommand{\UU}{{U}}
\newcommand{\TT}{{\bf G}}

\newcommand{\DD}{{D}}

\newcommand{\XI}{\Xi}

\renewcommand{\bA}{{\bf A}}
\renewcommand{\bH}{{\bf H}}
\newcommand{\bW}{{\bf W}}
\renewcommand{\bR}{{\bf R}}
\renewcommand{\bD}{{\bf D}}

\renewcommand{\PP}{\Pi}
\newcommand{\PPd}{\Pi^d}

\newtheorem{rem}{Remark}

\newtheorem{theorem}{Theorem}
\def\la{\langle}
\def\ra{\rangle}

\journal{Performance Evaluation}
\begin{document}
\begin{frontmatter}
\title{Global Optimal Solutions to  General
Sensor Network Localization Problem}

\author[first]{N. Ruan}
\ead{n.ruan@ballarat.edu.au}

\author[first,second]{D.Y. Gao\corref{cor1}}
\ead{d.gao@ballarat.edu.au, david.gao@anu.edu.au}

\cortext[cor1]{Corresponding author}

\address[first]{School of Sciences, Information Technology and Engineering,\\
Federation University  Australia, Mt Helen,  VIC 3353, Australia}
\address[second]{Research School of Engineering, Australian National University, Canberra,
 ACT 0200, Australia}

\begin{abstract}
Sensor network localization problem is to determine the position of
the sensor nodes in a network given pairwise distance
measurements. Such problem can be formulated as a
polynomial minimization via the least squares method.
This paper presents a canonical duality theory for
solving  this challenging problem.
It is shown that the nonconvex minimization problem can be reformulated as
a concave maximization dual problem over a convex set in a symmetrical matrix space,
and hence can be solved efficiently by combining
 a general (linear or  quadratic)  perturbation technique  with existing
optimization techniques.
Applications are illustrated by solving some relatively large-scale   problems.
Our results show that the general sensor network
localization problem is not NP-hard unless its canonical dual problem has no solution.
Fundamental ideas for solving general NP-hard problems are  discussed.
\end{abstract}
\begin{keyword}
Sensor network localization;\ Canonical duality theory;  Perturbation method;
Global optimization;\ NP-Hard problems
\end{keyword}
\end{frontmatter}
\section{Introduction}
Sensor network localization
is an important problem
in communication and information theory,
and   has attracted an increasing attention
\cite{bfadp10,dsam11,kkm09,rrh08,zhma08}.
The information collected through a sensor network can be interpreted and
relayed far more effectively if it is known where the
information is coming from and where
it needs to be sent. Therefore, it is often very useful to know the
positions of the sensor nodes in a network.
Wireless sensor network consists of a large number of
wireless sensors located in a geographical area with the ability to
communicate with their neighbors within a limited radio range.
Sensors collect the
local environmental information, such as temperature or humidity, and
can communicate with each other.
Wireless sensor network is applicable to a range
of monitoring applications in civil and military scenarios, such as geographical
monitoring, smart homes, industrial control and traffic monitoring.
There is  an urgent need to develop robust and efficient algorithms
that can identify sensor positions in a network by using
only the measurements of the mutual distances of the wireless sensors  from
their neighbors, which is called neighboring distance measurements.
The advance of wireless communication
technology has made the sensor network a low-cost and highly efficient
method for environmental observations.

Sensor network localization can also be formulated as an optimization
problem by least squares method.
However, this problem is nonconvex with many local minimizers.
To find  global optimal solutions by traditional theories and local-search methods  is fundamentally difficult.
It turns out that
the general sensor localization problem has been considered to be NP-hard \cite{agy,mw}.
Several approximation methods have been developed   for
solving this difficult optimization problem (see
 \cite{pal} and references cited therein).
The semi-definite programming  (SDP) and second-order
cone programming (SOCP) relaxations
are two of  the most popular methods studied recently
 \cite{biswas-lian-wang-ye,pots11,tseng,wzyb08}.
The basic idea of SDP relaxation is to think of the quadratic terms
as new variables subject to  a linear matrix inequality.
The SOCP relaxation is developed  in a similar way.
For both SDP and SOCP relaxation, computed
sensor locations are not accurate when the
solution of the
localization problem is not unique.  This is because  many numerical schemes,
such as primal-dual and interior point methods,
for SDP or SOCP relaxation often return to
the analytic centre of the solution set.
These solutions are, in general,
not global optimal solutions.

Mathematically speaking, the localization problem in $\real^d$ can be stated as
follows \cite{alf-00,chiang-amma}:
Consider a sensor network in $\real^d$ with $m$ anchors and $n$ sensors. An
anchor is a node whose location $a_k \in \real^d$, where
$k=1, \cdots, m$, is known,
and a sensor is a node whose location $x_i\in \real^d$,
where $i=1, \cdots,n$,  is
yet to be determined. For a pair of sensors
$x_i$ and $x_j$, their Euclidean distance is denoted as $d_{ij}$.
Similarly, for a pair of sensor $x_i$ and anchor $a_k$, their
Euclidean distance is denoted as $e_{ik}$. In general, not all pairs of sensor
/sensor and sensor/anchor are known, so the known pair-wise distances of
sensor/sensor and sensor/anchor are denoted as $(i,j) \in \calA_d$ and $(i,k)
\in \calA_e$, respectively.
However, if we directly apply the general least squares method, the
computation is very expensive and not practical for
large problems \cite{huang-chen-wang2011}.

Canonical duality theory developed from  nonconvex analysis
and global optimization  (see \cite{gao-book00, gawu11b})
is a powerful   methodology, which has been used
successfully for solving a large class of
challenging  problems in various disciplines. See, for example,
\cite{gao-cace09,gao-ruan-pardalos,Gao-Sherali-AMMA09,la-gao,jtb}.
This paper presents an effective perturbation  method based on
the canonical duality theory  to solve the general
sensor network localization problem.
Our  main contribution  is to show that this nonconvex optimization problem is not NP-hard
unless its canonical dual problem has no solution.
 The rest of this paper is organized as follows.
In the next section, we  first reformulate  the original problem as an optimization
problem, where the decision variable is expressed  in
tensor (matrix) forms. In Section 3,   the canonical duality theory is discussed in
matrix  space and a general
 analytical solution form  is
obtained by a complementary-dual principle.
In section 4, the
general sensor localization problem is first reformulated in vector space
and then transformed as a concave maximization dual problem
over  a convex feasible space $\calS^+_a$. Based on the triality theory,
a quadratic perturbation method is proposed, which shows that
the nonconvex  sensor network optimization problem is not NP-hard
unless its canonical dual problem has no solution in $\calS^+_a$.
 Section 5 presents  some concrete numerical experiments for sensor
localization problems with two,  18, 20
and 200 sensors. The cases
with noise are also considered. Results are compared with  standard semi-definite
programming method.
Concluding remarks are given in the last section.

The notations used in this paper are: $\real$ denotes the set of real numbers;
$A^T$denotes the transpose of matrix $A$. For a finite set $S$,
$|S|$ denotes its cardinality and the bilinear form
$\langle u,u^{\ast} \rangle $ is simply the scalar
product of two vectors or tensors.

\section{Problem Statement}
Let us consider a general sensor network localization problem, where
the sensor locations are to be determined by solving the system
of nonlinear equations
\begin{eqnarray}
(\calP_0): \;\;&&\|\bx_i- \bx_j\| = d_{ij},\  \; (i,j ) \in \calA_d, \label{po1}\\
\;\;\;\;\;\; &&\|\bx_i- \ba_k\| = e_{ik},\  \; ( i,k )\in \calA_e.\label{po2}
\end{eqnarray}
Here, the vectors  $\ba_k,\;k=1, \cdots, m $,  are specified anchors,
where
\begin{eqnarray*}
\| \bx_i  - \bx_j\| = \sqrt{ \sum_{\alp = 1}^d ( x_{i}^{\alp} - x_{j}^{\alp})^2 }
\end{eqnarray*}
denotes the Euclidian distance between locations $\bx_i$ and $\bx_j \in \real^d$, $i=1, \cdots,n; j=1,\cdots,n$,
and
\begin{eqnarray*}
\calA_d &=& \{ (i,j) \in [n] \times [n] \; | \;\;
\|\bx_i- \bx_j\| = d_{ij},\;   \;\;\; i < j, \;\; d_{ij}\; \mbox{are given distances}\},\\
\calA_e &=& \{ (i,k) \in [n] \times [m] \; | \;\;
\|\bx_i-\ba_k \|= e_{ik},\;  \;\;\;\; e_{ik}\; \mbox{are given distances}\},
\end{eqnarray*}
where $[N]=\{1,\cdots,N\}$  for any integer $N$.\\
For a small number of sensors, it might be possible to compute sensor
locations by solving equations (\ref{po1})-(\ref{po2}). However,
solving this algebraic system
can be very expensive computationally
when the number of sensors is large.

By the least squares method \cite{r-g-j},
the sensor network localization problem $(\calP_0)$ can be reformulated  as a
fourth-order polynomial optimization problem stated below.
\begin{eqnarray}
(\calP_1):\;\; & \min   &  \left\{  \PP(\bX) = 
\sum_{(i, j)\in \calA_d }\half w_{ij} (\|\bx_i -\bx_j\|^2-d_{i j}^2)^2 \right.\nonumber \\
&& \;\;\;\;\;\;\;\;\;\;\;\; \left.+ \sum_{(i, k)\in \calA_e }\half q_{ik} (\|\bx_i -\ba_k\|^2-e_{i k}^2)^2 \right\}, \label{eqanchor}
\end{eqnarray}
where $\bX=[x_1,x_2,\cdots, x_n]=
\{x_i^{\alpha}\}\in \real^{d\times n}$ is a  matrix with each column $x_i$ being
a position in $\real^d$, $w_{ij}, q_{i k}>0$ are given weights.
Obviously, $\bX$ are true sensor locations if and only if
the optimal value is zero.
This nonconvex optimization problem appears extensively in mathematical physics \cite{gao-yu},
chaotic dynamics \cite{ruan-gao-ima},
numerical algebra \cite{r-g-j},
computational biology \cite{jtb},
as well as  finite element analysis of structural mechanics
\cite{cai-gao-qin03,sato-gao}.
Due to the nonconvexity, this problem could have many local minimizers.
It is fundamentally difficult, or even impossible,
to find global optimal solutions by  traditional direct methods.
In the following, we shall see that by using the
canonical duality theory, this nonconvex minimization problem
can be reformulated as  a concave maximization dual problem over
 a convex set under certain conditions, which can be solved efficiently by
 a proposed perturbation method.

\section{Canonical duality theory: A brief review}
The canonical duality theory is composed mainly of 1) a canonical transformation; 2) a complementary-dual principle;
3) a triality theory.
This theory can be   demonstrated
by solving the following
general nonconvex problem (the primal problem $(\calP)$ in short)
\begin{eqnarray}
(\calP): \; \min_{ \xx \in \calX_a}
\left\{ \PP(\xx) = \half   \la \bx ,  \bA \xx  \ra  -
\la  \xx ,  \bff \ra  +  \WW( \bB \xx) \right\},
\end{eqnarray}
where $\calX_a \subset \real^{d\times n}$ is a
given  feasible space,  $\la \xx, \xx^* \ra $
denotes the bilinear form
between $\xx$ and its dual variable $\xx^*$,
$\bff \in \calX^*_a \subset \real^{n\times d}$ is a given matrix,
$ \bA :\calX_a \rightarrow \calX^*_a $ is  a given self-adjoint linear operator,
$\bB $ is a linear operator which assign each $\bx \in  \calX_a$  to  a (deformation gradient-like) variable
in a linear space $ \calW_a$, on which,
$\WW(\bw):\calW_a \rightarrow \real$ is a well-defined differentiable nonconvex function.

The   canonical   transformation
is to choose a ``geometrically admissible" nonlinear operator (see \cite{gao-book00})
\begin{eqnarray}
\bveps = \Lam (\bx):\calX_a \rightarrow \calE_a ,
\end{eqnarray}
which maps the convex set  $ \calX_a$ into a convex  set $\calE_a$,
and a {\em canonical function} $\VV: \calE_a \rightarrow \real$
such that the nonconvex functional $\WW( \bw)$
can be recast in  a canonical form
$\WW( \bB\bx) = \VV(\Lam(\bx))$.
Thus, the primal problem $(\calP)$  can be
written in the following canonical form:
\begin{eqnarray}
(\calP): \;  \min_{\bx \in \calX_a}
\left\{ \PP(\bx) =   \VV(\Lam(\bx)) - \UU(\bx)\right\},
\label{eq-canform}
\end{eqnarray}
where $\UU(\bx) =  \la \bx, \bff \ra - \half \la \bx , \bA \bx \ra$.
By the definitions introduced in
\cite{gao-book00}, a nonlinear operator $\Lam(\bx):\calX_a \rightarrow \calE_a$
is said to be geometrically admissible if it can be used as a (deformation) measure
such that the canonical transformation $W(\bB\bx) = V(\Lam(\bx))$ satisfies certain necessary
(geometrical and physical)  conditions, for examples,  the {\em   objectivity and isotropy.}
 Let
 \[
 {\cal R} = \{ \bR \in \real^{m\times m} | \; \bR^T = \bR^{-1}, \;\; \det \bR = 1 \}
 \]
 be a proper
 orthogonal rotation group in $\real^m$.
\begin{definition}[Objectivity and Isotropy \cite{gao-book00} ]

{\it   A subset $\calW_a  $ is said to be {\em objective} if
\eb
 \bR \bw \in \calW_a \;\; \forall \bw \in \calW_a   \;  \mbox{ and } \;  \forall \bR \in {\cal R}  .
 \ee
   A real-valued function $\WW:\calW_a  \rightarrow \real$ is said to be
 {\em objective}  if its domain is objective and
 \eb
\WW( \bR\bw) = \WW(\bw) \;\; \forall \bw \in \calW_a  \; \mbox{ and } \; \forall \bR \in {\cal R} .
 \ee

 A subset $\calW_a $ is said to be {\em isotropic} if
\eb
  \bw\bR\in \calW_a \;\; \forall \bw \in \calW_a   \;  \mbox{ and } \;  \forall \bR \in {\cal R} .
 \ee
 A real-valued function $\WW:\calW_a  \rightarrow \real$ is said to be
 {\em isotropic}  if its domain is isotropic and
 \eb
\WW( \bw\bR) = \WW(\bw) \;\; \forall \bw \in \calW_a  \; \mbox{ and } \; \forall \bR \in {\cal R}  .
 \ee
  }
 \end{definition}

Geometrically speaking, the objectivity  means that
the function $\WW (\bw)$ does not depend on  rotation, but only
on certain objective measure   of its variable $\bw$.
Therefore, the most simple objective function is the right Cauchy-Green  deformation  tensor
$\bC  = \bw^T \bw $ since
\[
\bC(\bR \bw) =  \bw^T \bR^T \bR \bw = \bw^T \bw = \bC(\bw) \succeq 0  \;\; \forall \bR \in {\cal R}.
\]
 While the isotropy implies  that the function $\WW(\bw)$ possesses a certain symmetry.
Clearly,  the left Cauchy-Green  deformation  tensor $\bw \bw^T$ is an isotropic measure due to the fact
 \[
 ( \bw\bR^T) (\bw \bR^T)^T  = \bw \bw^T   \succeq 0 \;\; \forall  \bR \in {\cal R}.
 \]
The concepts of objectivity and isotropy play  important role in Semi-Definite Programming (SDP) and
    integer programming  \cite{gao-jimo07,gao-ruan-jogo10}. Particularly, if $\bw$ is a vector,
    the objectivity is identical to isotropy.
    Furthermore, if the objective function $\WW(\bw)$ is considered as a  kinetic energy   and $\UU(\bx)$
    is viewed as the potential energy, then the function
    $\Pi(\bx)$ is the original  Lagrangian in mathematical physics \cite{gao-book00}.

The objectivity in science is also refereed as {\em frame invariance},   which
lays a foundation for mathematical physics and
systems theory.
In fact, the canonical duality theory was originally developed from this  concept
\cite{gao-book00}, which is the  reason  why this theory can be applied not only for modeling and
  analysis of complex systems, but also for solving a large class of nonconvex/nonsmooth/discrete
  problems in both mathematical physics and global optimization (see review article \cite{Gao-Sherali-AMMA09}).

A differentiable function $\VV(\bveps)$ is said to be
a \textit{canonical function}  on its domain  $\calE_a$ if the
duality mapping $\bvsig = \nabla \VV(\bveps)$ from $\calE_a$ to its range
$ \calE^*_a $
is invertible. Let   $\la \bveps ;  \bvsig \ra $
denote  the bilinear form on $\calE_a \times \calE^*_a$.
Thus, for the given canonical function $\VV(\bveps)$,
its Legendre conjugate
$\VV^*(\bvsig)$ can be defined uniquely by the Legendre transformation
\begin{eqnarray}
\VV^*(\bvsig)  = \sta \{ \la \bveps ;  \bvsig \ra - \VV(\bveps ) \; | \; \; \bveps \in \calE_a \},
\end{eqnarray}
where  the notation $\sta \{ g(\bveps) | \; \bveps \in \calE_a\}$
stands for finding stationary point of $g(\bveps)$ on $\calE_a$.
It is easy to prove that
the following canonical  duality relations hold on $ \calE_a \times \calE^*_a$:
\begin{eqnarray}
\bvsig =\nabla \VV(\bveps) \; \Leftrightarrow \;  \bveps = \nabla
\VV^*(\bvsig) \; \Leftrightarrow
\VV(\bveps) + \VV^*(\bvsig) =
\la \bveps ; \bvsig \ra  . \label{eq-candual}
\end{eqnarray}
By this one-to-one canonical duality,   the nonconvex term
$W(\bD \bx)=\VV(\Lam(\bx))$ in the problem $(\calP)$ can be replaced by
$\la \Lam(\bx) ; \bvsig \ra -
\VV^*(\bvsig)$ such that the nonconvex
function $\PP(\bx)$ is reformulated  as
the so-called Gao and Strang total complementary function \cite{gao-book00}:
\begin{eqnarray}
\Xi(\bx, \bvsig) = \la  \Lam(\bx) ; \bvsig \ra    - \VV^*(\bvsig) -  \UU(\bx).
\label{eq:xi}
\end{eqnarray}
By using this total complementary function,
the canonical dual function $\PP^d (\bvsig)$
can be obtained   as
\begin{eqnarray}
\PP^d(\bvsig) &=& \sta \{ \Xi(\bx, \bvsig) \; | \; \bx \in \calX_a   \} \nonumber \\
&=& \UU^\Lam(\bvsig) - \VV^*(\bvsig), \label{pidg}
\end{eqnarray}
where    $\UU^\Lam(\bvsig)$ is defined by
\begin{eqnarray}
\UU^\Lam(\bvsig) =  \sta \{ \la  \Lam(\bx) ;  \bvsig \ra - \UU(\bx) \;
| \;\; \bx \in \calX_a \}. \label{ulam}
\end{eqnarray}
In many applications, the geometrically nonlinear operator
$\Lam(\bx)$ is usually  a tensor-valued quadratic function
\begin{eqnarray}
\Lam(\bx)= \left\{ \half \la \bx, \bH_{kl} \bx \ra   \right\} : \real^{d\times n} \rightarrow \real^{n\times n},
\end{eqnarray}
where $\bH_{kl}:\calX_a \rightarrow \calX^*_a  \;\; \forall k, l = 1, \dots, n  $
 is a symmetrical linear operator.
In this case, the canonical dual variable $\bvsig \in \calE^*_a \subset
\real^{n\times n}$ is a symmetrical tensor and
 the total complementary function $\Xi$ can be written in the following form
\eb
\Xi(\bx, \bvsig) = \half \la \bx , \bG(\bvsig) \bx \ra -   \VV^*(\bvsig) - \la \bx, \bff \ra,
\ee
where
\eb
\bG(\bvsig) = \bA+\sum_{k,l} \vsig_{kl} \bH_{kl}. \label{eq-ggenral}
 \ee
 For any given $\bvsig \in \calE^*_a$, the criticality condition $\nabla_\bx \XI(\bx, \bvsig) = 0$
 leads to the canonical equilibrium equation
  $\bG(\bvsig) \bx = \bff$. Let
\eb
\calS_a = \{ \bvsig \in \calE^*_a | \;  \det  \bG(\bvsig) \neq  0   \} .
\ee
Then on $\calS_a$,   the solution to canonical equilibrium equation
can be written as $\bx = \bG^{-1}(\bvsig) \bff$.
Therefore,
  replacing the primal variable $\bx$ by  this generalized solution in $\Xi$, the canonical dual function
  (\ref{pidg}) can be explicitly formulated in the form of
\begin{eqnarray}
\PPd(\bvsig)=-\half \la \bff, \bG^{-1} (\bvsig) \bff \ra -V^{\ast}(\bvsig).
\end{eqnarray}

\begin{theorem}[Complementary-Dual Principle \cite{gao-book00}]\label{duality}
The   function $\Pi^d(\bvsig)$ is canonically dual to $\Pi(\bx)$ in the sense that
if $\barbvsig$ is a critical point of $\Pi^d(\bvsig)$, then
\begin{eqnarray}\label{eq-anasol}
\barbx = \bG^{-1}(\barbvsig) \bff
\end{eqnarray}
is a critical point of $\Pi(\bx)$ and
\begin{eqnarray}
\PP(\barbx) =  \Xi(\barbx, \barbvsig) = \PP^d(\barbvsig). \label{eq-p=d}
\end{eqnarray}
Conversely, if $\barbx$ is a solution to $(\calP)$, it must be in the form of
(\ref{eq-anasol}) for critical solution $\barbvsig$ of $\Pi^d(\bvsig)$.
\end{theorem}

This theorem has extensive applications in nonconvex
analysis and global optimization \cite{gao-cace09}.
In finite deformation theory, this complementary-dual principle solved a 50-years open problem
  \cite{li-gupta}. Note that the  feasible set $\calS_a$ is not convex,
 in  order to identity the extremality property of the critical solutions,
 we need to introduce the following subsets of $\calS_a$:
\[
\calS^+_a = \{\bvsig\in \calS_a|\; G(\bvsig) \succ  0\}, \;\;
 \calS^-_a = \{\bvsig\in \calS_a|\; G(\bvsig) \prec 0\}.
\]
 \begin{theorem}[Triality Theory]
Suppose the
$(\barbx, \barbvsig)$ is a critical point of $\Xi(\bx,\bvsig)$.
The critical solution $\barbx  $
is a unique global minimizer of $(\calP) $ if and only if
 $\barbvsig \in \calS^+_a$ is a global maximizer of $\PP^d(\bvsig)$ on $\calS^+_a$, i.e.
\begin{eqnarray}
\PP(\barbx) =  \min_{\bx \in \calXa } \PP(\bx)  \; \Leftrightarrow \;\;
\max_{\bvsig \in \calS_a^+} \PP^d(\bvsig) = \PP^d(\barbvsig) \label{trisaddle}.
\end{eqnarray}
 If $\barbvsig \in \calS^-_a$, then $\barbvsig $ is a local maximizer of $\PP^d(\bvsig)$ on its
 neighborhood $\calS_o \subset \calS^-_a$
 if and only if
$\barbx  $
is a local maximizer of $(\calP) $ on its  neighborhood $\calX_o \subset \calX_a$, i.e.
\begin{eqnarray}
\PP(\barbx) =  \max_{\bx \in \calX_o } \PP(\bx)  \;\; \Leftrightarrow \;\;
\max_{\bvsig \in \calS_o} \PP^d(\bvsig) = \PP^d(\barbvsig)  \label{trimax}.
\end{eqnarray}

If $\barbvsig \in \calS^-_a$ and $\dim \calX_a = \dim \calS_a$, then $\barbvsig $ is a local minimizer of $\PP^d(\bvsig)$ on its
 neighborhood $\calS_o \subset \calS^-_a$
 if and only if
$\barbx  $
is a local minimizer of $(\calP) $ on its  neighborhood $\calX_o \subset \calX_a$, i.e.
\begin{eqnarray}
\PP(\barbx) =  \min_{\bx \in \calX_o } \PP(\bx)  \;\; \Leftrightarrow \;\;
\min_{\bvsig \in \calS_o} \PP^d(\bvsig) = \PP^d(\barbvsig) \label{trimin}.
\end{eqnarray}
\end{theorem}

\begin{rem}
The saddle min-max duality theorem (\ref{trisaddle})
was first proved by Gao and Strang in finite deformation theory \cite{GaoStrang89},
while the double-min and double-max duality statements were discovered in 1996.

The double-max duality statement (\ref{trimax}) can be proved easily by the fact that
\[
 \max_{\bx \in \calX_o} \max_{\bvsig \in \calS_o} \Xi(\bx, \bvsig)
= \max_{\bvsig \in \calS_o} \max_{\bx\in \calX_o} \Xi(\bx, \bvsig)
\;\;\forall (\bx ,\bvsig) \in \calX_o \times \calS_o \subset \calX_a \times \calS^-_a.
\]

The  double-min duality statement  (\ref{trimin}) holds only under the condition $\dim \calX_a = \dim \calS_a$,
which was an open problem discovered in 2003 \cite{gao-opt03,gao-amma03} and solved recently
  in \cite{gawu11b}.  If $\dim \calX_a \neq  \dim \calS_a$, this double-min duality holds in a weak form (see \cite{gawu11b}).

Based on the triality theory, the nonconvex minimization problem $(\calP)$ is equivalent
(only if $ \calS_a^+ \neq \emptyset $)
to a concave maximization dual problem
\eb
(\calP^d): \;\;
\max \left\{ \PP^d(\bvsig) \; | \; {\bvsig \in \calS_a^+}  \right\} \label{cd0}.
\ee
Although $\Pi^d(\bvsig)$ contains an inverse matrix of $\bG(\bvsig)$,
this canonical dual problem can be solved easily by some well-developed
nonlinear optimization techniques (see \cite{fang-gaoetal07,wang-etal})
as long as $\calS^+_a$ contains at least one critical point of $\Pi^d(\bsig)$.
Otherwise, a perturbation method will be discussed in the next section.

\end{rem}

\noindent {\bf Example 1 (Fundamental idea of linear perturbation)}.

\begin{figure}[h]
\begin{picture}(3,2)
\setlength{\unitlength}{.5cm}
\put(0,-10){{\large \psfig{file=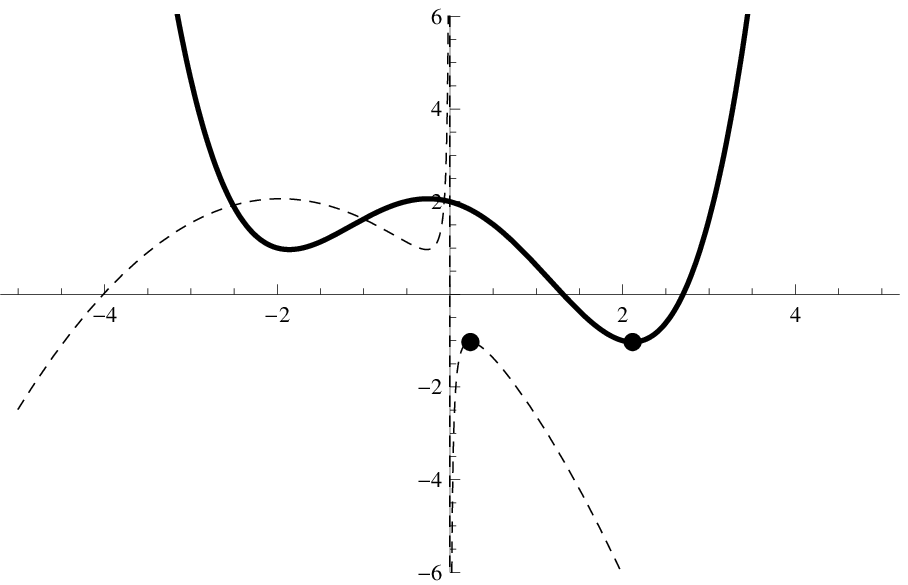,height=4cm, width =6cm}}}
\put(14,-10){{\large \psfig{file=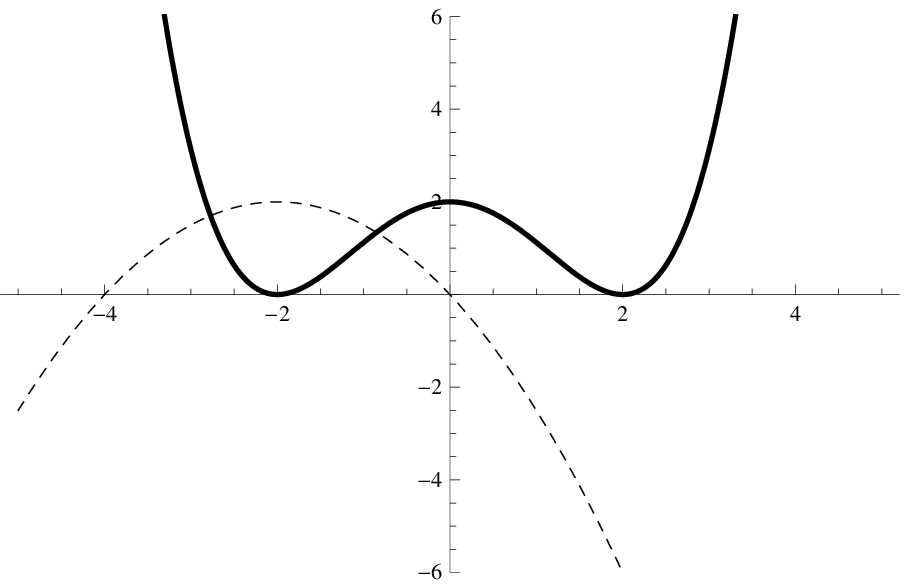,height=4cm, width =6cm}}}
\put(4,-12){(a) $f=0.5$}
\put(16,-12){(b)   $f = 0$}
\end{picture}\vspace{6.0cm}
\caption{Graphs of the primal function $\Pi(\bx)$  (solid)
and its canonical dual  $\Pi^d(\vsig)$ (dashed). }\label{onedim}
\end{figure}

To  explain the theory, let us  consider a very simple nonconvex optimization in
$\real^n$:
\begin{eqnarray}
\min \left\{ \Pi(\bx)=\half \alpha \left(\half\|\bx\|^2-\lam \right)^2-\bx^T \bff  \; \;\; \forall \bx \in \real^n \right\} ,
\end{eqnarray}
where $\alp, \lam > 0$ are given parameters.
The criticality condition $\nabla P(\bx)=0$ leads to a nonlinear algebraic
equation system in $\real^n$
\begin{eqnarray}
\alpha (\half \|\bx\|^2-\lam)\bx =\bff.
\end{eqnarray}
Clearly, to solve this nonlinear algebraic equation directly   is difficult.
Also traditional convex optimization theory
 can not be used to identify global minimizer.  However, by the
canonical dual transformation, this problem can be solved completely and easily.
To do so, we let  $\xi=\Lam(\bx)=\half\|\bx\|^2  \in \real$, which is an objective measure. Then,
the nonconvex function $W(\bx) = \half \alpha(\half \| \bx \|^2 -\lam)^2$
can be written in canonical form
$V(\xi) = \half \alpha (\xi - \lam)^2$.
Its Legendre conjugate is  given by
$V^{\ast}(\vsig)=\half \alpha^{-1}\vsig^2 + \lam \vsig$, which is strictly convex.
Thus,
the total  complementary function for this nonconvex optimization
problem is
\begin{eqnarray}
\Xi(\bx,\vsig)= \half \|\bx\|^2   \vsig-\half
\alpha^{-1}\vsig^2 - \lam \vsig - \bx^T \bff.
\end{eqnarray}
For a fixed $\vsig \in \real$, the  criticality condition
$\nabla_{\bx} \Xi(\bx)=0$ leads to
\begin{eqnarray}\label{balance}
\vsig \bx-\bff=0.
\end{eqnarray}
For each  $\vsig \neq 0  $,
the  equation
(\ref{balance}) gives $\bx=\bff/\vsig$ in vector form. Substituting this into the
total complementary function $\Xi$,
the canonical dual function can be easily obtained as
\begin{eqnarray}
\Pi^d(\vsig)&=&\{\Xi(\bx,\vsig)| \nabla_{\bx} \Xi(\bx,\vsig)
=0\}\nonumber\\
&=& -\frac{\bff^T \bff}{2 \vsig}-\half \alpha^{-1} \vsig^2
-\lam \vsig, \;\;\; \forall \vsig\neq 0.
\end{eqnarray}
The critical point of this canonical function is obtained
by solving the following dual algebraic  equation
\begin{eqnarray}
(\alpha^{-1} \vsig+\lam)\vsig^2=\half \bff^T \bff. \label{eq-deuler}
\end{eqnarray}
For any given parameters $\alpha$, $\lam$ and the vector $\bff\in \real^n$,
this cubic algebraic equation has at most three real  roots
satisfying $\vsig_1 \ge 0 \ge \vsig_2\ge \vsig_3$,
and each of these roots leads to a critical point of the nonconvex function
$P(\bx)$, i.e., $\bx_i=\bff/\vsig_i$, $i=1,2,3$. By
the fact that $\vsig_ 1 \in \calS^+_a = \{ \vsig \in \real\; |\; \vsig > 0 \}$,
$\vsig_{2,3} \in \calS^-_a = \{ \vsig \in \real\; |\; \vsig < 0 \}$,
then Theorem 2 tells us that $\bx_1$ is
a global minimizer of $\Pi(\bx)$, $\bx_3 $ is a local maximizer of $\Pi(\bx)$,
while $\bx_2$ is a local minimizer if $n=1$ (see Fig. 1).
If we choose    $n = 1, \;\; \alpha= 1$, $\lam=2$, and $f= \half$,
the primal function and canonical dual function
are shown in Fig. \ref{onedim} (a), where,  $x_1= 2.11491$ is global minimizer
of $\Pi(\bx)$, $\vsig_1=0.236417$ is global maximizer of $\Pi^d(\bvsig)$, and
$\Pi(x_1)=-1.02951=\Pi^d(\vsig_1)$ (see the two black dots).
Also it is easy to verify that $x_2 $ is a local minimizer, while $x_3$ is a local maximizer.

If we let $\bff= 0$, the graph of $\Pi(\bx)$ is symmetric (i.e. the so-called double-well potential or the Mexican hat for $n=2$
\cite{gao-amma03})
 with  infinite number of global minimizers
 satisfying $\| \bx \|^2  = 2 \lam$.
In this case, the canonical dual $\Pi^d (\vsig) = - \half \alp^{-1} \vsig^2 - \lam \vsig$ is strictly concave
with only one critical point (local maximizer) $\vsig_3 = -  \alp \lam  \in \calS_a^-$ (for $\alp, \lam > 0$).
The corresponding solution $\bx_3 = \bff /\vsig_3 = 0$ is a local maximizer.
 By the canonical dual equation (\ref{eq-deuler})
we have $\vsig_1 = \vsig_2 = 0$ located on the boundary of $\calS^+_a$,
which  corresponding to the two global minimizers   $x_{1,2} = \pm \sqrt{2 \lam}$ for $n=1$, see Fig. 1 (b).
This is exactly the example of one sensor $\bx = (x, 0) \in \real^2$
and two anchors $\ba_{1,2} = (0, \pm a)$ with $e_1 = e_2 = b$. 
Due to symmetry $(f = 0)$, the problem $(\calP)$ has two possible solutions  $\bx_{1,2} =  ( x_{1,2}, 0)$ with
$\lam = \half (b^2 - a^2)$.

This simple example shows a fundament issue in global optimization, i.e.,
the  optimal solutions of a nonconvex problem depends sensitively on the linear term (input) $\bff$.
Geometrically speaking, the objective function $\WW(\bB \bx)$
in $\Pi(\bx)$ possesses certain symmetry.
If there is no linear term  (subjective function) in $\Pi(\bx)$,
 the nonconvex problem usually has  more than one
global minimizer due to the symmetry.
Traditional direct approaches and the popular SDP method are usually failed to deal with this situation.
By the canonical duality theory, we understand that in this case the canonical dual function has no critical point in
 $\calS^+_a$.
Therefore,  by  adding a linear perturbation $\bff$ to destroy
 this  symmetry,  the canonical duality theory can be used to solve
the nonconvex problems to obtain one of global optimal solutions.
This idea was originally from Gao's work (1996) on post-buckling analysis of large deformed beam \cite{gao-mrc96}, where the triality theorem was first proposed  \cite{gao-amr}. The potential energy of this beam model is a double-well function, similar to this sensor example,
  without lateral force or imperfection, the beam could have two buckling states
(corresponding to two minimizers) and one un-buckled state (local maximizer).
Later on (2008) in the Gao and Ogden  work on analytical solutions in phase transformation \cite{gao-ogden},
 they further discovered that the nonconvex system has no phase transition unless the  force distribution
  $f(x)$ vanished at
certain points.
 They also discovered that if force field $f(x) $ changes dramatically, all the Newton type direct approaches
failed  even to find any  local minimizer.
The linear perturbation method has been used successfully for solving global optimization problems \cite{chen-gao-oms,r-g-j,silva-gao,wzyb08}.

\section{Application to Sensor Network Localization Problem}
Now let us tern our attention for solving the general sensor network optimization
problem $(\calP_1)$.
For convenience,  we transfer variables from
matrix to vectors, and let
\begin{eqnarray*}
\by &=& [x_{1}^{1} \cdots x_{1}^d  \cdots  x_{n}^{ 1}  \cdots  x_{n}^{ d}]^T
\in \real^{nd}: \mbox{Locations of sensors\; (variables)},\\
\bW &=& [w_{1 1}  \cdots w_{1 n} \cdots  w_{n 1}
\cdots  w_{n n}]^T \in \real^{n n}: \\
&&\mbox{Weights for the optimization problem}\;(\calP_1),\\
\bQ &=& [q_{1 1}  \cdots q_{1 m} \cdots  q_{n 1}
\cdots  q_{n m}]^T \in \real^{n m}:\\
&&\mbox{Weights for the optimization problem}\; (\calP_1),\\
\ba &=& [\sum_{\alpha=1}^d(a_{ 1}^{ \alpha})^2, \cdots,
\sum_{\alpha=1}^d(a_m^{\alpha})^2]^T:\\
&& \mbox{Sums of squares of  anchors},\\
\bd &=& [d_{1 1}^2 \cdots  d_{1 n}^2 \cdots  d_{n 1}^2  \cdots d_{n n}^2]^T \in
\real^{n n}: \mbox{Squares of distances between sensors},\\
\be &=& [e_{1 1}^2 \cdots  e_{1 m}^2 \cdots  e_{n 1}^2  \cdots e_{n m}^2]^T \in
\real^{n m}:\\
&&\mbox{Squares of distances between sensors and anchors}.
\end{eqnarray*}
Then, Problem $(\calP_1)$ can be written in a vector form  given below.
\begin{eqnarray*}
(\calP): \;\; \min
 & & \left\{
\PP(\by) = \sum_{(i,j)\in A_d}
\half w_{ij}\left(\by^T D_{ij}\by- d_{ij}^2 \right)^2\right.\\
&&\left.+\sum_{(i,k)\in A_e}
\half q_{ik}\left(\by^T E_{ik} \by
-2 A_{ik}^T \by +\sum_{\alpha=1}^d (a_{ik}^{\alpha})^2- e_{ik}^2 \right)^2
\right\},
\end{eqnarray*}
where $A_{ik}$    and $a_{ik}^\alp$  are components of the anchors  $\ba_k$ obtained from the
expansion of $\| \bx_i - \ba_k \|^2$ in equation (\ref{eqanchor}),
$E_{ik}\in \real^{nd \times nd}$ is a diagonal matrix defined by
\begin{eqnarray*}
E_{ik}=\left[
\begin{array}{ccc}
0&0&0\\
0&I_{ik}&0\\
0&0&0
\end{array}
\right],
\end{eqnarray*}
with $I_{ik}\in \real^{d \times d}$
being the identity matrix corresponding to sensor $i$ and anchor $ k$,
so that the (1,1) entry of $I_{ik}$ coincides with
the $(i,k)$ entry of $E_{ik}$. Similarly,
$D_{ij}$ is an $nd \times nd$ matrix  defined by
\begin{eqnarray*}
D_{ij}=\left[
\begin{array}{ccccc}
0&0&0&0&0\\
0&I_{ii}&0& -I_{ij}&0\\
0&0&0&0&0\\
0&-I_{ji}&0&I_{jj}&0\\
0&0&0&0&0
\end{array}
\right],
\end{eqnarray*}
with $I_{ii}$,$I_{jj}$, $I_{ij}$,$I_{ji}$ $\in \real^{d \times d}$
being the identity  matrices, so that the (1,1) entry of $I_{ii}$
coincides with the $(i,i)$ entry of the matrix $D_{ij}$.
For $I_{jj}$, $I_{ij}$, $I_{ji}$, they are defined similarly.  Let
\begin{eqnarray}
\xi_{ij}=\Lam_{ij}(\by) &=&\by^T\DD_{ij} \by,\\
\eps_{ik} =\Lam_{ik}(\by) &=&\by^T E_{ik} \by- 2 A_{ik}^T\by,
\end{eqnarray}
where $\Lam_{ij}$ and $\Lam_{ik}$ are, respectively, geometrical operators
from $ \real^{nd}$ into
\begin{eqnarray*}
\calE_{d} &=& \{ \bveps \in  \real^{nn} |\; \veps_{ij} \ge 0,
\; \veps_{ij}  =0  \; \mbox{if}\; i=j \}
\end{eqnarray*}
and
\begin{eqnarray*}
\calE_{e} &=& \{ \beps \in  \real^{mn} |\; \eps_{ik}\ge 0\}.
\end{eqnarray*}
By introducing  quadratic functions $V_{\xi}: \calE_{d} \rightarrow\real $
and $V_{\eps}: \calE_{e} \rightarrow\real $ such that
\begin{eqnarray}
V_{\xi} (\xi_{ij}) &=&  \half \sum_{(i,j) \in \calA_d}  w_{ij} (\xi_{ij}-d_{ij}^2)^2
\end{eqnarray}
and
\begin{eqnarray}
V_{\eps}  (\eps_{ik} ) = \half  \sum_{(i,k) \in  \calA_e}
 q_{ik}  \left(\eps_{ik}
+\sum_{\alpha=1}^d(a_{ik}^{\alpha})^2-e_{ik}^2 \right)^2
\end{eqnarray}
Problem $(\calP)$ can  then be reformulated in the
canonical form given below:
\begin{eqnarray*}
(\calP): \;\;  \min  \left\{  \Pi(\by)= \VV_{\xi}(\Lam_{ij}(\by))
+ \VV_{\eps}(\Lam_{ik}(\by)) |
\;\; \by \in \real^{nd} \right\} .
\end{eqnarray*}
Note that the  function $\VV_{\xi}(\xi_{ij})$
and $\VV_{\eps}(\eps_{ik})$ are both
 convex. Then the following  duality relations are invertible
\begin{eqnarray}
\vsig_{ij} = \nabla  \VV_{\xi}(\xi_{ij})= w_{ij} (\xi_{ij}-d_{ij}^2)
,\;\;(i,j) \in \calA_d,
\end{eqnarray}
and
\begin{eqnarray}
\sig_{ik} = \nabla  \VV_{\eps}(\eps_{ik})= q_{ik} \left(\eps_{ik}
+\sum_{\alpha=1}^d(a_{ik}^{\alpha})^2-e_{ik}^2 \right)
,\;\;(i,k) \in \calA_e,
\end{eqnarray}
where $\vsig_{ij}$ and $\sig_{ik}$ are dual variables.
Let ${\cal S}_d$ be the range of the  duality mapping
 $\vsig_{ij} =
\nabla V_{\xi}(\xi_{ij})$,
and let ${\cal S}_e$ be the range of the  duality mapping
$\sig_{ik} =
\nabla V_{\eps}(\eps_{ik})$.
Then, for any given $\bvsig \in {\cal S}_d$ and $\bsig \in {\cal S}_e$,
the Legendre conjugate $V_{\xi}^*$ and $V_{\eps}^*$ can be uniquely defined by
\begin{eqnarray}
V_{\xi}^* (\vsig_{ij} ) &=& {\rm sta}\left\{ \sum_{(i,j) \in \calA_d} \xi_{ij}  \vsig_{ij} -V_{\xi} (\veps_{ij} ) \; | \;
\veps_{ij} \in {\cal V}_d\right\} \nonumber \\
&=&
 \sum_{(i,j) \in \calA_d}  \left( \half  w_{ij}^{-1} \vsig_{ij}^2+ d_{ij}^2 \vsig_{ij} \right)
\end{eqnarray}
and
\begin{eqnarray}
V_{\eps}^* (\sig_{ik} ) &=& {\rm sta}\left\{  \sum_{(i,k) \in \calA_e}  \eps_{ik}  \sig_{ik}
 -V_{\eps} (\eps_{ik} ) \; | \; \eps_{ik} \in {\cal V}_e\right\}  \nonumber \\
&=& \sum_{(i,k) \in \calA_e}  \left[ \half q_{ik}^{-1} \sig_{ik}^2
 + \left(e_{ik}^2-\sum_{\alpha=1}^d(a_{ik}^{\alpha})^2 \right)\sig_{ik} \right].
\end{eqnarray}
Clearly, $(\bxi, \bvsig)$ and $(\beps, \bsig)$ form a {\em canonical duality pair}
(see \cite{gao-book00}).
The following canonical duality relations hold on
both ${\calE}_d \times {\cal S}_d$ and
${\calE}_e \times {\cal S}_e$
\begin{eqnarray*}
\bvsig   &=& \nabla \VV_{\xi}(\bveps )  \;\; \Leftrightarrow  \;\;
\bveps  = \nabla \VV_{\xi}^*(\bvsig )
\Leftrightarrow \la \bveps ; \bvsig  \ra = \VV_{\xi}(\bveps ) + \VV_{\xi}^*(\bvsig ),\\
\bsig   &=& \nabla \VV_{\eps}(\beps )  \;\; \Leftrightarrow  \;\;
\beps  = \nabla \VV_{\eps}^*(\bsig )
\Leftrightarrow  \la \beps ; \bsig \ra  = \VV_{\eps}(\beps ) + \VV_{\eps}^*(\bsig ),
\end{eqnarray*}
respectively.
So the generalized complementary
function (\cite{gao-book00}) is given by
\begin{eqnarray}
\Xi (\by, \bvsig,\bsig)
&=&   \sum_{(i,j) \in A_d} \Lam_{ij}(\by)
\vsig_{ij}   -  V_{\xi}^*(\bvsig ) +
\sum_{(i,k) \in A_e} \Lam_{ik}(\by) \sig_{ik}   - V_{\eps}^*(\bsig ) \nonumber\\
&=& \half \by^T \bG(\bvsig,\bsig) \by- \bF^{T}(\bsig) \by
-\half (\bW^{-1} )^T (\bvsig \circ \bvsig)\nonumber\\
&&-\half (\bQ^{-1})^T (\bsig \circ \bsig)
-\bd^T \bvsig+\ba^T \bsig -\be^T \bsig,\label{comple1}
\end{eqnarray}
where $\bs \circ \bt := [s_1 t_1, s_2 t_2,\cdots, s_n t_n]^T$ denotes the Hadamard
product of any two vectors $\bs$, $\bt \in \real^n$,
\begin{eqnarray}
\bF(\bsig) =
\left[
\sum_{k=1}^m 2 a_k^1 \sig_{1k}
\cdots
\sum_{k=1}^m 2 a_k^d \sig_{1k}
\cdots
\sum_{k=1}^m 2 a_k^1 \sig_{nk}
\cdots
\sum_{k=1}^m 2 a_k^d \sig_{nk}
\right]^T,
\end{eqnarray}
\begin{eqnarray}
\bG(\bvsig, \bsig)= 2(\Diag(F_1(\bvsig))
+\Diag(F_2(\bsig))
+ G_3(\bvsig)),
\end{eqnarray}
with
\begin{eqnarray*}
F_1(\bvsig) =
\left[
\begin{array}{c}
\sum_{i=1}^n \vsig_{1i}\\
\vdots\\
\sum_{i=1}^n \vsig_{1i}\\
\vdots\\
\sum_{i=1}^n \vsig_{ni}\\
\vdots\\
\sum_{i=1}^n \vsig_{ni}
\end{array}
\right],\;
F_2(\bsig) =
\left[
\begin{array}{c}
\sum_{k=1}^m \sig_{1k}\\
\vdots\\
\sum_{k=1}^m \sig_{1k}\\
\vdots\\
\sum_{k=1}^m \sig_{nk}\\
\vdots\\
\sum_{k=1}^m \sig_{nk}
\end{array}
\right],
\end{eqnarray*}
\begin{eqnarray*}
G_3(\bvsig) =
\left[
\begin{array}{ccc}
-\vsig_{11}I_{11}&\cdots&-\vsig_{1n}I_{1n}\\
\vdots&\vdots&\vdots\\
-\vsig_{n1}I_{n1}&\cdots&-\vsig_{nn} I_{nn}
\end{array}
\right].
\end{eqnarray*}
where the notation $\Diag(F_1)$ represents a
diagonal matrix with $F_{1i}$, $i=1, \cdots, n$
being its diagonal elements. For a fixed $\bvsig \in \calS_d$ and $\bsig \in \calS_e$,
the criticality condition $\nabla_{\by}
\Xi(\by,\bvsig,\bsig)=0$ leads to the following
{\em  canonical equilibrium equation}:
\begin{eqnarray}\label{equi}
\bG(\bvsig,\bsig) \by -\bF(\bsig)=0.
\end{eqnarray}
Substitute the solution of this equation into (\ref{comple1}),
the canonical dual function can be formulated as:
\begin{eqnarray*}
\PPd(\bvsig,\bsig)  &=&  -\half \bF(\bsig)^T \TT^{-1} (\bvsig,\bsig) \bF(\bsig)
 -\half (\bW^{-1} )^T (\bvsig \circ \bvsig)\\
&&-\half (\bQ^{-1})^T (\bsig \circ \bsig)
-\bd^T \bvsig+\ba^T \bsig -\be^T \bsig,
\end{eqnarray*}
where
\begin{eqnarray}
\bF(\bsig) =
\left[
\sum_{k=1}^m 2 a_k^1 \sig_{1k}
\cdots
\sum_{k=1}^m 2 a_k^d \sig_{1k}
\cdots
\sum_{k=1}^m 2 a_k^1 \sig_{nk}
\cdots
\sum_{k=1}^m 2 a_k^d \sig_{nk}
\right]^T,
\end{eqnarray}
and
\begin{eqnarray*}
\bW^{-1} &=& \left[\frac{1}{w_{1 1}}  \cdots  \frac{1}{w_{1 n}}
\cdots  \frac{1}{w_{n 1}}  \cdots  \frac{1}{w_{n n}}
\right]^T,\\
\bQ^{-1} &=& \left[\frac{1}{q_{1 1}}  \cdots  \frac{1}{q_{1 m}}
\cdots  \frac{1}{q_{n 1}}  \cdots  \frac{1}{q_{n m}} \right]^T.
\end{eqnarray*}
Therefore, the canonical dual problem can be written in the form given below.
\begin{eqnarray*}
(\calP^{d}): \;\;
 \sta   \left\{
\PP^{d}(\bvsig, \bsig)   
 | \;\; \bvsig \in \calS_d, \bsig \in \calS_e
 \right\} .
\end{eqnarray*}
By Theorem \ref{duality}, we have following results:
\begin{theorem}\label{dualitysdp}
Problem $(\calP^d)$ is
a canonical dual of  the primal problem $(\calP)$ in the sense that if
 $(\barbvsig, \barbsig)$ is a critical point of $({\mathcal P}^d)$, then
\begin{eqnarray}\label{xformu}
\barby=\TT ^{-1} (\barbvsig, \barbsig)\bF(\barbsig)
\end{eqnarray}
is a critical point of $({\mathcal P})$ and
\begin{eqnarray}
\PP (\barby)= \PP^d(\barbvsig, \barbsig).
\end{eqnarray}
\end{theorem}

Theorem \ref{dualitysdp}   shows that the nonconvex primal problem $(\calP)$ is
equivalent to its canonical dual problem $(\calP^d)$ with zero duality gap, and
the  solution of   $(\calP)$
can be analytically expressed by (\ref{xformu}) in terms of the canonical dual variables.
The global minimizer can be identified by the saddle min-max duality theorem (\ref{trisaddle}).
In this case,
 the  feasible space $\calS^+_a$ should be
\begin{eqnarray}
&&\calS_a^+=\{(\bvsig, \bsig) \in \calS_d \times \calS_e \; | \;\; \TT (\bvsig, \bsig) \succ 0\}.
\end{eqnarray}
Based on the triality theory,  the nonconvex  primal  problem
$(\calP)$ is equivalent to the following canonical dual problem:
\begin{eqnarray}
(\calP^d_{\max}): \;\;
\max \{ \PP^d(\bvsig,\bsig) |  \;\; (\bvsig,\bsig) \in \calS_a^+ \}.
\end{eqnarray}

\begin{theorem}\label{opcr}
If  $(\barbvsig, \barbsig) \in \calS^+_a$ is a critical point of the canonical dual function
$\PP^d(\barbvsig, \barbsig)$,
then it is a unique solution of $(\calP^d_{\max})$,
the vector  $\barby = \bG^{-1}(\barbvsig, \barbsig)F(\barbsig) $
is a unique global optimal solution to the nonconvex sensor network optimization problem
$(\calP)$, and
\begin{eqnarray}\label{maxi}
\PP(\barby)=\min_{\by\in {\real^{nd}}} \PP(\by) = \max_{(\bvsig,\bsig)\in \calS_a^+}
\PP^d(\bvsig,\bsig)=\PP^d(\barbvsig,\barbsig).
\end{eqnarray}
\end{theorem}

{\bf Proof}. First, we need to prove the convexity of $\calS^+_a$.
We let $\bxi^* = ( \bvsig, \bsig)$.
For any given $\bxi^*_1, \bxi^*_2 \in \calS^+_a$, we should have
\[
\theta \bG(\bxi^*_1) \succ 0, \;\; (1- \theta ) \bG(\bxi^*_2) \succ 0 \;\; \forall \theta \in [0, 1].
\]
Therefore,
\[
\theta \bG(\bxi^*_1)  + (1- \theta ) \bG(\bxi^*_2) = \bG(\theta \bxi^*_1 + (1- \theta) \bxi^*_2) \succ 0
\;\;\forall \theta \in [0, 1].
\]
This shows that $\calS^+_a$ is convex.

By the fact that the total complementary function $\Xi(\by, \bxi^*)$ is a saddle function on $\real^{nd} \times \calS^+_a$,
the classical saddle min-max duality  theory (cf. \cite{eke-tem}, page 57, or \cite{gao-book00}, page 39) leads to
\[
\min \Pi(\by) = \min_{\by \in \real^{nd} } \max_{\bxi^* \in \calS^+_a} \Xi(\by, \bxi^*)
= \max_{\bxi^*  \in \calS^+_a} \min_{\by \in \real^{nd} } \Xi(\by, \bxi^* ) = \max_{\bxi^*  \in \calS^+_a} \Pi^d(\bxi^*).
\]
Therefore, by the complementary-dual principle, the critical point $\barby   \in \real^{nd} $  of
$\Xi$ is a
global min of $\Pi(\by)$ if and only if the associated critical point $\bar{\bxi^*} $ is a global max of $\Pi^d(\bxi^*)$ on
$\calS^+_a$.
Since $\Xi(\by, \bxi^*)$ is strictly convex in $\by$ and concave in $\bxi^*$  on  $\real^{nd}  \times \calS^+_a$,
its saddle point is unique.
  \hfill $\Box$\\

\begin{rem}[NP-Hard Problems and Perturbations]
It is known that  the sensor localization problem is
   NP-hard only in the worst case (see \cite{agy}).
However, what is the worst case is not clear to general  problems. By the
canonical duality theory we know  that for a large class of nonconvex  and integer optimization
problems, as long as their canonical dual problems or perturbed forms have
critical points in the dual feasible domain $S_a^+$, these  challenging  problems can
be solved easily by convex optimization techniques (see \cite{gao-cace09}). Otherwise, these
  problems could be  NP-hard, which is a conjecture proposed by Gao in 2007 \cite{gao-jimo07}.

By the fact that  $\Pi^d(\bvsig)$ is concave on the  open set  $\calS^+_a$,
  the canonical dual problem $(\calP^d)$ may have no critical point in $\calS^+_a$.
It is our understanding that  a NP-hard  optimization problem  usually possesses
 certain symmetry in its  modeling if the primal problem has more than one global minimal solution.
The main idea of the linear perturbation is to destroy this symmetry such that the associated
 perturbed problem has a unique solution. This method is well-known in engineering mechanics
and  was   introduced  to global optimization by the authors in 2008
 for  solving   quadratic equations via least squares method
\cite{r-g-j}. Recently, this method has been used successfully to solve a NP-complete max-cut problem
  \cite{wang-etal} as well as some challenging problems in global optimization (see \cite{chen-gao-oms,morales-gao-opt}).
For complex systems, how to correctly chose the linear perturbation vector $\bdelta$ is still an open problem.
Therefore, some  high-order  perturbation methods proposed in 2010 for solving certain NP-hard nonconvex/integer
optimization problems \cite{gao-ruan-jogo10}.
Particularly,  a {\em quadratic perturbation method } can be suggested as the following:
\eb
  \min_{\by} \max_{(\bvsig, \bsig) \in \calS^+_{\mu_k}} \left\{
 \Xi_k(\by, \bvsig, \bsig) = \Xi(\by, \bvsig, \bsig) + \half \rho_k \| \by - \by_k \|^2
- \la  \by, \bdelta \ra \right\}  \label{eq-perminmax}
\ee
where $\rho_k > 0$ is a given parameter, $\by_k \in \real^{nd}$ is a given vector (previous solution in iteration process),   $\bdelta \in \real^{nd}$ is a given linear perturbation vector,
$\calS^+_{\mu_k}$ is a  relaxed canonical dual feasible space defined by
\eb
\calS^+_{\mu_k} = \{ (\bvsig, \bsig) \in \calS_d \times \calS_e | \;\;
\bG(\bvsig, \bsig) + \mu_k \bI \succeq 0 \},
\ee
where $0< \mu_k < \rho_k$ is given relaxation parameter.

By the fact that
the perturbed total complementary function
$\Xi_k(\by, \bvsig, \bsig)$ is strictly convex in $\by \in \real^{nd}$ and strictly concave
in $(\bvsig, \bsig)$ on the closed convex set $\calS^+_{\mu_k}$,
an effective  canonical primal-dual algorithm can be
developed for solving the
saddle min-max problem (\ref{eq-perminmax}).
\end{rem}

\section{Numerical Simulations}
In this section, we will first look at a simple  case of a  network with  four anchors and two
sensors. The locations of the anchors are known while the locations of the
sensors are to be determined. The linear perturbation method
will be  used to show how the symmetry can be destroyed such that  this small scale network localization problem
can be solved nicely. We
then move on to formulate more general sensor networks   by  randomly generated test problems.
The so-called root mean square distance (RMSD) will be used  to measure
the accuracy of the estimated positions.
\subsection{A four-anchor sensor network localization problem}
\begin{figure}[!t]
\centering
\includegraphics[width=2.5in]{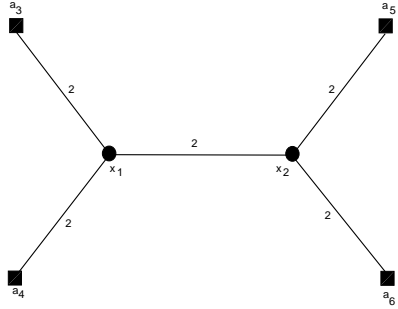}
\caption{Sensor network with two sensors and four anchors.}
\label{fig-sensor2}
\end{figure}

Consider the sensor network problem with  two sensors and four anchors,
as shown in Fig \ref{fig-sensor2}.
Let $\bx_1 = [ x_1^1, x_1^2]^T,\;\bx_2 =
[x_2^1, x_2^2 ]^T \in \real^2$ denote the locations of the unknown sensors
and let $ \ba_i,  \; i=3,4,5,6$, denote the locations of the four known anchors.
The  sensor network location problem
is to  solve the following  system of nonlinear  equations:
\begin{eqnarray*}
(\calP_0)\;\;\;\;&&\|\bx_1- \bx_2\| = d_{12},\; \|\bx_1- \ba_3\| = e_{13},
\;\|\bx_1- \ba_4\| = e_{14},\\
&&\|\bx_2- \ba_5\| = e_{25},\;
\|\bx_2- \ba_6\| = e_{26},
\end{eqnarray*}
where
\begin{eqnarray*}
&&d_{12 } = e_{13} = e_{14}= e_{25}=e_{26}=2, \\
&&\ba_3 = [ -2, \sqrt{3}= 1.7321]^T,\;\; \ba_4 = [ -2, -\sqrt{3}= -1.7321]^T,\\
&&\ba_5 = [ 2, \sqrt{3}= 1.7321]^T,\;\; \ba_6 = [ 2, -\sqrt{3}= -1.7321]^T.
\end{eqnarray*}
Let  $\by=[x_1^1 \;  x_1^2  \; x_2^1 \;  x_2^2]^T \in \real^4$.

Then, the fourth-order polynomial $\PP(\by)$ is given by
\begin{eqnarray*}
\PP(\by) &=&\half \left((x_1^1-x_2^1)^2+(x_1^2-x_2^2)^2-2^2 \right)^2
+  \half((x_1^1+2)^2+(x_1^2-\sqrt{3})^2-2^2)^2  \\
&+ & \half((x_1^1+2)^2+(x_1^2+\sqrt{3})^2-2^2)^2
+  \half((x_2^1-2)^2+(x_2^2-\sqrt{3})^2-2^2)^2 \\
& + & \half((x_2^1-2)^2+(x_2^2+\sqrt{3})^2-2^2)^2.
\end{eqnarray*}
\normalsize
To solve this symmetrical sensor network problem  $(\calP_0)$,
we  use a linear perturbation  method (see \cite{r-g-j})
\begin{eqnarray}
(\calP_\delta): \;\; \min \left \{ \PP_\delta(\by)
= \PP(\by) -\bdelta^T \by \;  |   \;\;  \by \in \real^4 \right\},
\end{eqnarray}
where $\bdelta=[\delta_1^1 \; \delta_1^2 \; \delta_2^1 \; \delta_2^2]^T \ge 0 $
is a given perturbation vector.
On the canonical dual feasible space $\calS_a$ defined by
\begin{eqnarray}
\calS_a=\{(\vsig_{12}, \sig_{13}, \sig_{14}, \sig_{25}, \sig_{26})^T
\; |\; \vsig_{12}+\sig_{13}+\sig_{14} \ne 0,
\vsig_{12}+\sig_{25}+\sig_{26} \ne 0\},
\end{eqnarray}
the canonical dual problem to the $\delta$-perturbed problem $(\calP_{\delta})$ is
\begin{eqnarray}
(\calP^d_\delta): \;\; \max \left\{
\PP^d_\delta(\bvsig,\bsig )    |     \;  (\bvsig,\bsig) \in \calS_a\right\},
\end{eqnarray}
where
\begin{eqnarray*}
\PP^d_\delta(\bvsig,\bsig )
&=&
-\half F_\delta(\bsig)^T \TT^{-1}(\bvsig,\bsig) F_\delta (\bsig)
-(d_{12})^2 \vsig_{12}+(\ba_3^T \ba_3
-(d_{13})^2)\sig_{13}\\
&&+ (\ba_4^T \ba_4-(d_{14})^2)\sig_{14}+(\ba_5^T \ba_5- (d_{25})^2)\sig_{25} \\
&&  +(\ba_6^T \ba_6- (d_{26})^2)\sig_{26}
- \half \vsig_{12}^2-\half \sig_{13}^2-\half \sig_{14}^2
-\half \sig_{25}^2-\half \sig_{26}^2,
\end{eqnarray*}
\begin{eqnarray}
F_{\delta}(\bsig) &=&
( \delta_1^1+2 a_3^1 \sig_{13}+2 a_4^1 \sig_{14},
\delta_1^2+2 a_3^2 \sig_{13}+2 a_4^2 \sig_{14},\nonumber \\
&&\delta_2^1+2 a_5^1 \sig_{25}+2 a_6^1 \sig_{26},
\delta_2^2+2 a_5^2 \sig_{25}+2 a_6^2 \sig_{26}
)^T,
\end{eqnarray}
\begin{eqnarray*}
& & \TT(\bvsig,\bsig)=\\
& & {\small \left[
\begin{array}{cccc}
2(\vsig_{12}+\sig_{13}+\sig_{14}) & 0 & -2 \vsig_{12} & 0\\
0& 2(\vsig_{12}+\sig_{13}+\sig_{14}) & 0& -2 \vsig_{12} \\
-2 \vsig_{12} & 0& 2(\vsig_{12}+\sig_{25}+\sig_{26}) & 0\\
0 & -2 \vsig_{12} &0 & 2(\vsig_{12} + \sig_{25}+\sig_{26})
\end{array}
\right].}
\end{eqnarray*}
Set $\bdelta=[0.005, 0.005, 0.005, 0.005]^T$. Then,
the canonical dual  problem   has a unique solution \cite{gao-ruan-jogo10}
\begin{eqnarray*}
(\barbvsig,\barbsig) &=& [\vsig_{12}, \sig_{13}, \sig_{14}, \sig_{25}, \sig_{26}]^T\\
&=& [-0.0000, 0.0005, 0.0020, -0.0020, -0.0005]^T .
\end{eqnarray*}
By Theorem \ref{opcr},  it follows that
\begin{eqnarray*}
\barby &= &[ \barx_1^1, \barx_1^2, \barx_2^1, \barx_2^2 ]^T
=\TT^{-1}(\barbvsig,\barbsig) \FF_{\delta} (\barbsig) \\
 &=&  [-0.9994, 0.0002, 1.0006, 0.0002]^T
\end{eqnarray*}
\normalsize
is  a global minimizer of $\PP_\delta(\by)$.

It is easy to verify that
\begin{eqnarray*}
\Pi_\delta (\barby)=-4.1667\times 10^{-6}= \Pi^d_\delta (\barbvsig,\barbsig).
\end{eqnarray*}
By the fact that
\begin{eqnarray*}
\PP(\barby) = 4.1667\times 10^{-6},
\end{eqnarray*}
the $\delta$-perturbed solutions $\barby$ can be considered as
the global minimizer to  the original problem $(\calP_0)$ and
we have
\begin{eqnarray*}
&&\|\bx_1 -\bx_2\| = 2.0000,\;
\|\bx_1 -\ba_3\|  = 2.0001,\;
\|\bx_1 -\ba_4\|  = 2.0005,\\
&&\|\bx_2 -\ba_5\|  = 1.9995,\;
\|\bx_2 -\ba_6\| = 1.9999.
\end{eqnarray*}\\

\subsection{18  sensors network localization problem with four anchors}
We now consider sensor network localization problem with 18 sensors.
In this case, we have Problem $(\calP_1)$ with  $d=2$. Define
$\by= [ x_1^1, x_1^2, \cdots, x_n^1, x_n^2]^T \in \real^{2n}$, and let $w_{ij}
= q_{ik}=1$ in Problem $(\calP_1)$. Here,   we do not consider noise.

The  18 sensors
$\{\hat{\bx}_i=[\hat{x}_i^1,\hat{x}_i^2]: i=1, \cdots,18 \}$
are randomly generated in the unit square
[-0.5, 0.5] $\times$ [-0.5, 0.5].
The four anchors
$(\ba_1,\ba_2,\ba_3,\ba_4)$ are placed at the positions
$(\pm 0.45, \pm 0.45)$.
The distances
$\bd=\{d_{ij}\}$, $i=1,\cdots,18$; $j=1,\cdots,18$,
and $\be=\{ e_{ik} \}$,
$i=1,\cdots,18$; $k=1,\cdots,4$, are computed as follows:
\begin{eqnarray*}
d_{ij}=\|x_i^*-x_j^*\|,\;
e_{ik}=\|x_i^*-a_k\|
\end{eqnarray*}
We now assume that the locations of the 18 sensors are unknown. They are
to be determined by
the approach proposed in the paper.
The sequential quadratic programming approximation with active set strategy in the
optimization toolbox within the Matlab environment is used to solve the
canonical dual problem.

By Theorem \ref{opcr}, we obtain $\barby=[\barbx_1,\cdots,\barbx_{18}]^T$
with $\barbx_i=[\barbx_i^1,\barbx_i^2]^T, i=1, \cdots,18 $, which
is  a global minimizer of $\PP(\by)$.

\begin{figure}[!t]
\centering
\includegraphics[width=3.5in]{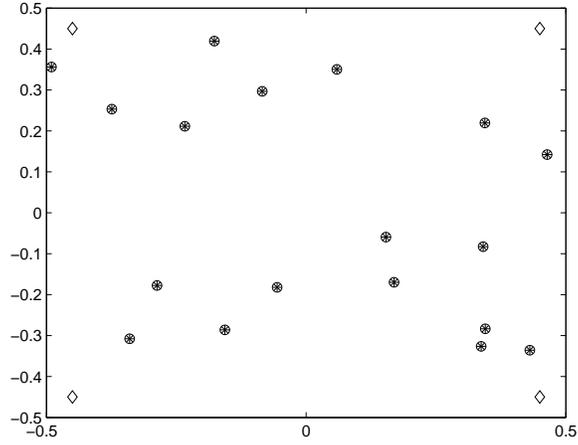}
\caption{Sensor network with 18 sensors by the canonical dual method.}
\label{fig-sensor18dual}
\end{figure}

\begin{figure}[!t]
\centering
\includegraphics[width=3.5in]{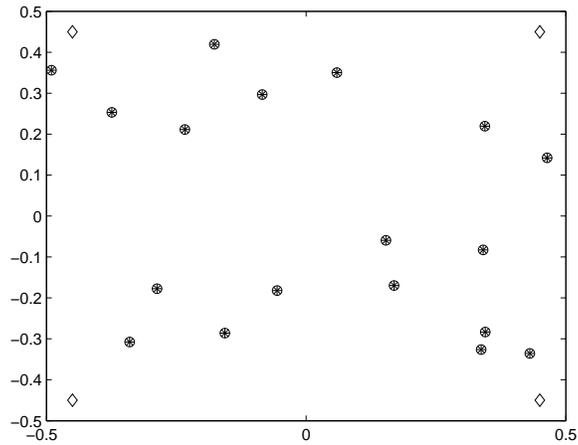}
\caption{Sensor network with 18 sensors by the standard SDP method.}
\label{fig-sensor18SDP}
\end{figure}

Furthermore, we have
\begin{eqnarray*}
\Pi(\barby)=1.30 \times 10^{-8} \simeq 3.03 \times 10^{-8}
= \Pi^d (\barbvsig,\barbsig).
\end{eqnarray*}
This problem is also solved by the standard semi-definite programming (SDP) method.
The RMSD obtained using the canonical dual method is $4.61 \times 10^{-7}$,
while the RMSD obtained using the
standard SDP method is $4.45 \times 10^{-5}$, where RMSD is the
Root Mean Square Distance defined by
\begin{eqnarray*}
\mbox{RMSD} =(\frac{1}{n} \sum_{i=1}^n \| \hat{\bx}_i- \barbx_i\|^2)^{\half},
\end{eqnarray*}
which is to measure the accuracy of the computed locations.\\
The computed results by the canonical dual method and the standard SDP method are
plotted in Fig. \ref{fig-sensor18dual} and Fig. \ref{fig-sensor18SDP},
respectively. The true sensor locations (denoted by circles) and
the computed locations (denoted by stars) are connected by solid lines.
Our program is implemented in the  MATLAB environment, where
SEDUMI \cite{lu09} is used as the SDP solver.
From the results obtained,  we see that, when there is  no noise and the sensor size is not
too large, both
the canonical dual method and SDP method are  very
effective method for finding sensor locations. In particular,
for the canonical dual method,
all the stars are exactly located inside circles.
The computational time for canonical dual method and SDP method
are 0.61 seconds and 21.53 seconds, respectively.
\subsection{A 20-sensor-network localization problem with distance errors}
A network of 20 uniform randomly distributed unknown points is generated in the
square area $[0,1]\times [0,1]$. We assume:
\begin{eqnarray*}
&&\mbox{if}\;\;\|x_i-x_j\| \leq  \mbox {radio range,
a distance (with noise) is given between}\\
&& x_i\;\; \mbox{and}\;\; x_j ;\\
&&\mbox{if}\;\; \|x_i-x_j\| >  \mbox {radio range,
no distance is given between}\;\; x_i\;\; \mbox{and}\;\; x_j.
\end{eqnarray*}
Also, there are four  anchors are located  in [0,0], [0,1], [1,0] and
[1,1]. The distances between the nodes are  calculated. If the distance between two
nodes is within the specified radio range of 0.4, the distance is included in the
edge set for solving the problem after adding a random error to it in the
following manner:
\begin{eqnarray*}
d_{ij}=\hat{d}_{ij}|1+N(0,0.001)|
\end{eqnarray*}
where $\hat{d}_{ij}$ is the actual distance between the 2 nodes,
and $N(0,0.001)$ is a random variable.

\begin{figure}[!t]
\centering
\includegraphics[width=3.5in]{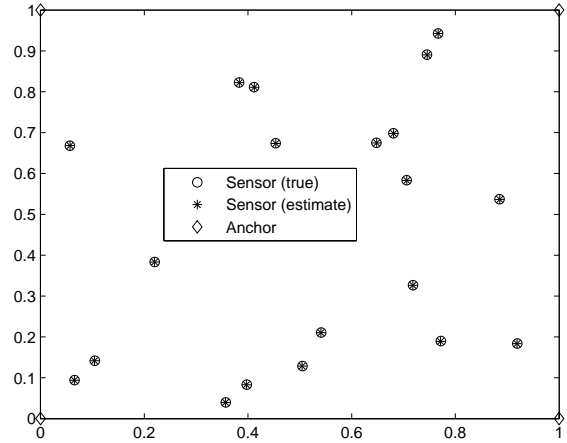}
\caption{Sensor network with 20 sensors solved by the canonical dual method.}
\label{fig-sensor20dual}
\end{figure}

\begin{figure}[!t]
\centering
\includegraphics[width=3.5in]{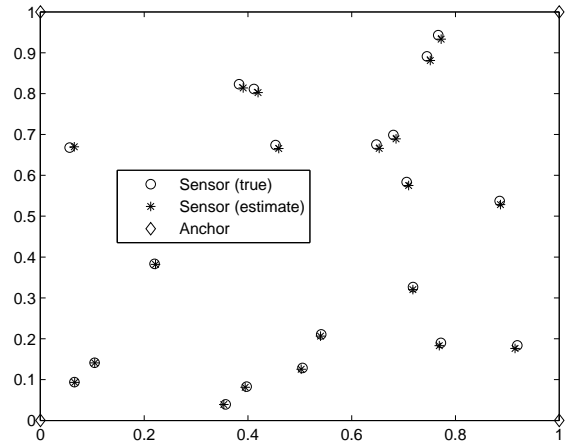}
\caption{Sensor network with 20 sensors solved by the standard SDP method.}
\label{fig-sensor20SDP}
\end{figure}

The computed results obtained by the canonical dual method and the standard SDP
method \cite{kim-kojima-waki} are
plotted in Fig. \ref{fig-sensor20dual} and Fig. \ref{fig-sensor20SDP}, respectively.
The true sensor locations (denoted by circles) and
the computed locations
(denoted by stars) are connected by solid lines. The computational time for
canonical dual method and SDP method are 0.65 seconds and 27.71 seconds.
\subsection{A 200-sensor-network localization problem with distance errors}
A network of 200 uniform randomly distributed unknown points is generated in the
square area $[0,1]\times [0,1]$. Four  anchors are located in [0,0], [0,1], [1,0] and
[1,1]. For all sensors, the radio range = 0.3.
The distance, including a random error, is generated in the
following manner:
\begin{eqnarray*}
d_{ij}=\hat{d}_{ij}|1+N(0,0.001)|
\end{eqnarray*}
\begin{figure}[!t]
\centering
\includegraphics[width=3.5in]{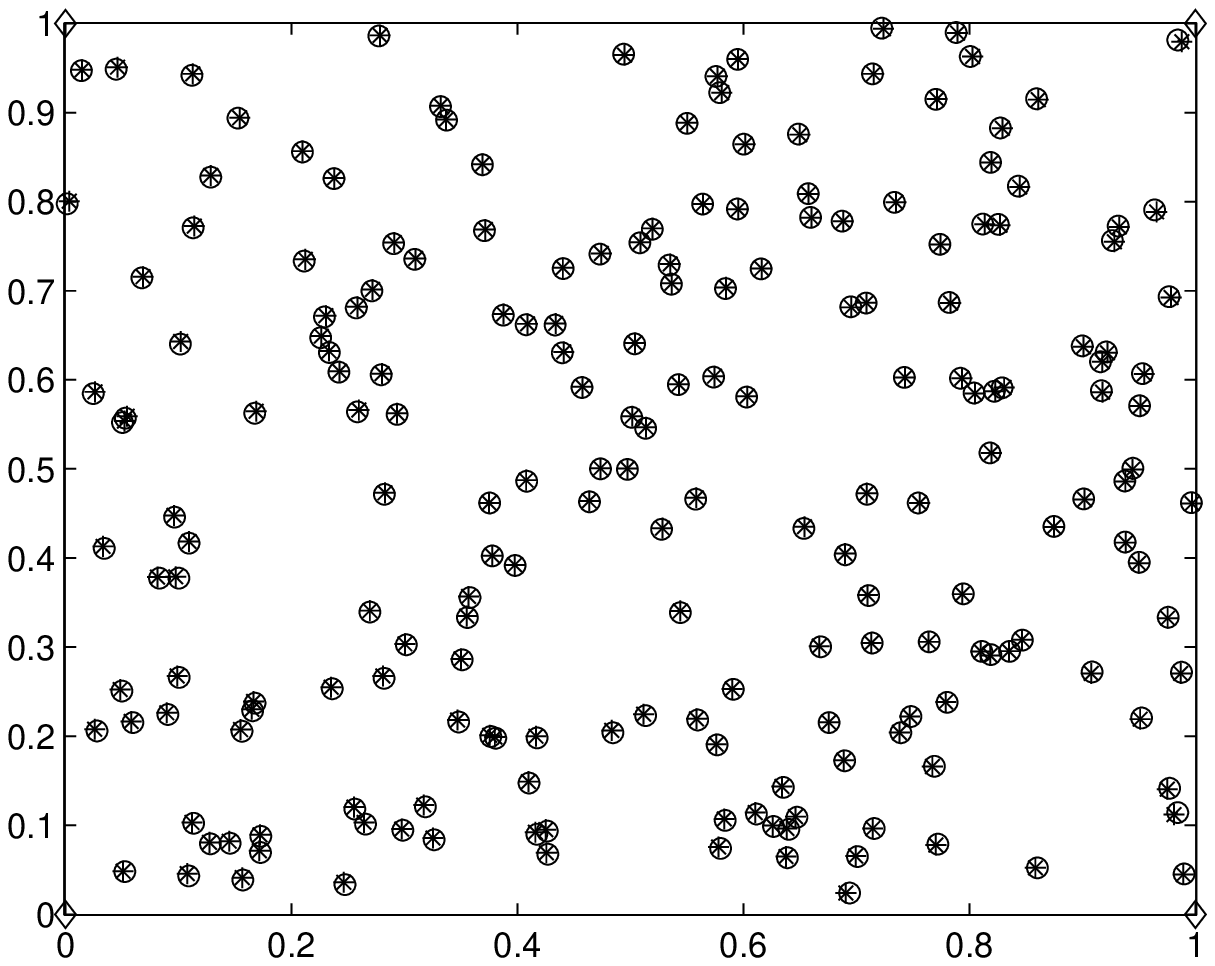}
\caption{Sensor network with 200 sensors solved by the canonical dual method.}
\label{fig-sensor200dual}
\end{figure}
\begin{figure}[!t]
\centering
\includegraphics[width=3.5in]{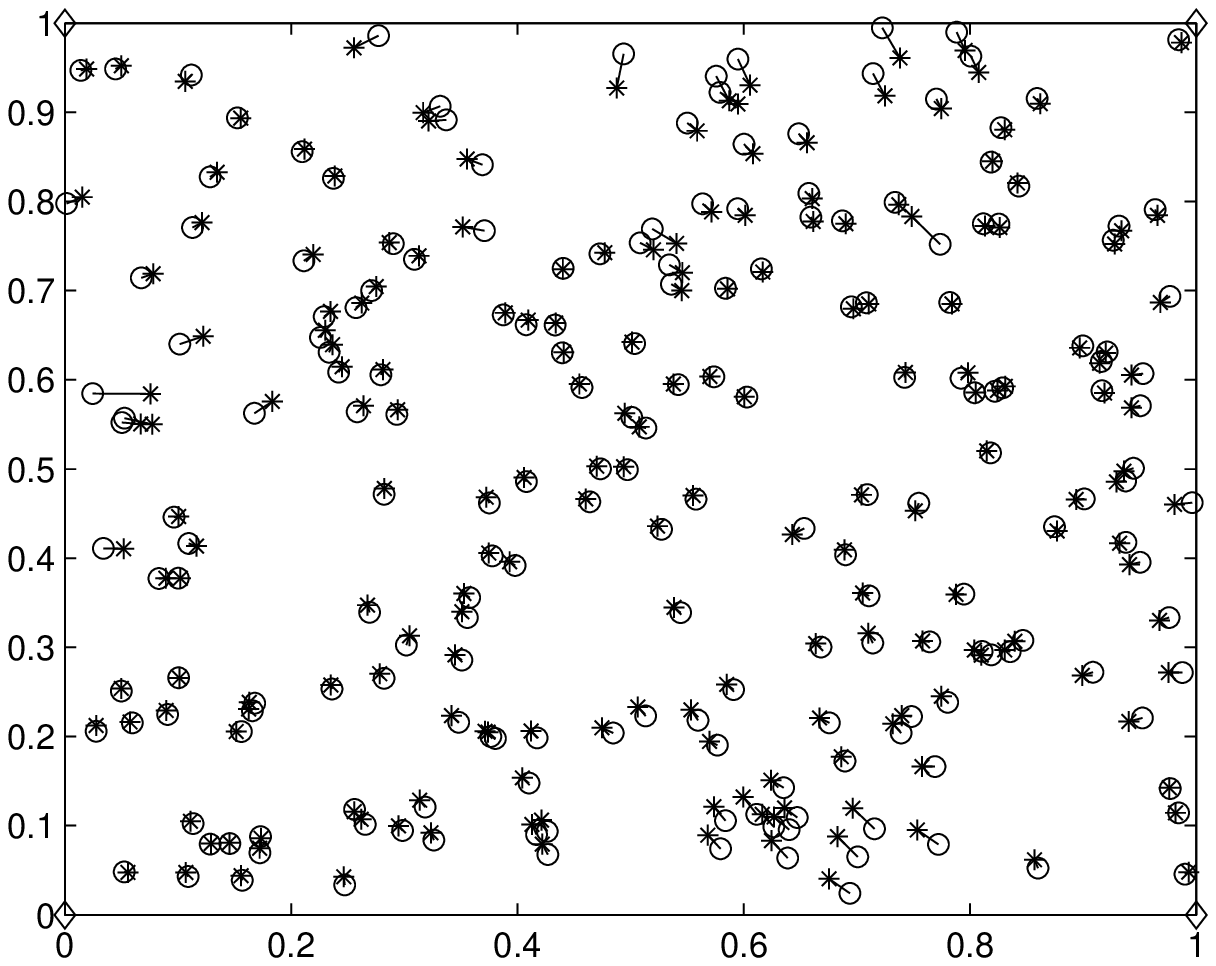}
\caption{Sensor network with 200 sensors solved by the standard SDP method.}
\label{fig-sensor200SDP}
\end{figure}
The computed results obtained by the quadratic perturbed canonical dual method and the
standard SDP method are
plotted in Fig. \ref{fig-sensor20SDP} and Fig. \ref{fig-sensor200SDP}, respectively.
Careful examination of the results obtained for
the cases involving  20 sensors  and 200 sensors, we observe that when  noise
is taken into consideration,
the canonical dual method gives rise to  much better solutions.
In particular, if the level of noise  or the sensor size is large,
the standard SDP is usually having difficulty to
finding  the exact sensor positions, see Figure \ref{fig-sensor20SDP} and
Figure \ref{fig-sensor200SDP}. The computational time for canonical
dual method and SDP method are 127.10 seconds and 1088.70 seconds, respectively.

\section{Conclusion Remarks}
We have presented a solid  application of the   canonical duality
theory for solving a general  sensor network localization problem.
By using the complementary-dual principle, a general form  of    analytical solution form is
obtained in terms of the canonical dual variables.
 Based on a  perturbation method,
  an effective canonical saddle min-max approach is proposed for solving this challenging problem.
  Our results show that the general sensor localization  problem  is not NP-hard if its canonical dual or perturbed
  problem  has a solution in $\calS^+_a$.
  Applications are illustrated by  detailed analysis of a small size problem as well as some
 relatively large scale  sensor network localization problems.

 From mechanics point of view, a  sensor network is similar to a structure,
 for the given boundary conditions (anchors) and external force $\bff$ (linear perturbation),
 as long as the problem is statically determinate,
  i.e. the canonical equilibrium equation (\ref{equi}) has a solution
   (see page 199 \cite{gao-book00}), the canonical dual problem has a critical point  in $\calS^+_a$
   and the sensor location      problem can be solved efficiently regardless of its size.
    By the definition (\ref{eq-ggenral}) of $\bG(\bvsig)$ we know that $\calS^+_a \neq \emptyset$
    as long as there exists at least one $\bH_{kl}  \succ 0$ (this condition satisfied naturally for any objective measure $\Lam(\bx)$).
  The necessary condition  has been studied
   in \cite{gao-cace09} (Theorem 8) and \cite{gao-ruan-jogo10} (Section 4) for  $\calS^+_a$ has a critical point of $\Pi^d(\bvsig)$.
   However, the sufficient condition is still open,
   which is fundamentally important for using linear perturbation method to solve  NP-hard problems.
  Nevertheless, the quadratic perturbation methods introduced in \cite{gao-ruan-jogo10} as well as in this paper
  (\ref{eq-perminmax}) provide more robust approach for solving this type of challenging problems.
  Finally, the performance of the method proposed on problems with
noisy distance data needs further investigation.\\
$\;$\\
{\bf Acknowledgment}:
The authors are gratefully indebted with the handling editor for his detailed remarks and important suggestions.
 This paper has   benefited greatly from two  anonymous referees' comprehensive and
  constructive comments.
  The main ideas and results of this paper have been announces at the
19th International Conference on Neural Information Processing (ICONIP2012), Nov. 12-15, 2012, Doha, Qatar
as well as several tutorial/plenary lectures in 2013 including the 3rd World Congress of Global Optimization,
July 7-12, 2013, the Yellow Mountains, China.
This work  was partially supported by a grant (AFOSR FA9550-10-1-0487)
from the US Air Force Office of Scientific Research. Dr. Ning Ruan was
supported by a funding from the Australian Government
under the Collaborative Research Networks (CRN) program.


\end{document}